# Speeding up explicit numerical evaluation methods for micromagnetic simulations using demagnetizing field polynomial extrapolation

Serban Lepadatu[*]

*Jeremiah Horrocks Institute for Mathematics, Physics and Astronomy, University of Central Lancashire, Preston PR1 2HE, U.K.*

**Abstract**

The performance of numerical micromagnetic models is limited by the demagnetizing field computation, which typically accounts for the majority of the computation time. For magnetization dynamics simulations explicit evaluation methods are in common use. Higher order methods call for evaluation of all effective field terms, including the demagnetizing field, at all sub-steps. Here a general method of speeding up such explicit evaluation methods is discussed, by skipping the demagnetizing field computation at sub-steps, and instead approximating it using polynomial extrapolation based on stored previous exact computations. This approach is tested for a large number of explicit evaluation methods, both adaptive and fixed time-step, ranging from $2^{nd}$ order up to $5^{th}$ order. The polynomial approximation order should be matched to the evaluation method order. In this case we show higher order methods with polynomial extrapolation are more accurate than lower order methods with full evaluation of the demagnetizing field. Moreover, for higher order methods we show it is possible to achieve a factor of 2 or more computation speedup with no decrease in solution accuracy.

[*] SLepadatu@uclan.ac.uk

# I. Introduction

Numerical micromagnetic modelling based on the Landau-Lifshitz-Gilbert (LLG) [1] equation is an essential tool for magnetization dynamics studies, including magnetization switching of nanoparticles [2], thin films [3], spin-valves and other multi-layered structures [4,5], domain wall motion studies [6-8], spin-torque-induced skyrmion dynamics [9-11], spin-waves [12] and spin-torque nano-oscillator simulations [13]. Explicit evaluation methods are in common use for solving the LLG equation. As is common practice, with higher order methods the effective field terms entering the LLG equation are evaluated at all sub-steps, including the computationally costly demagnetizing field. Indeed, all finite-difference micromagnetic modelling to date employs this approach where explicit evaluation methods are used, for example as implemented in OOMMF [14], Mumax3 [15], Fidimag [16], micromagnum [17], and Grace [18]. In Boris [19] this is also the default approach. It should also be noted that implicit evaluation methods are also available for micromagnetics [20-22], and these are particularly useful for stiff problems [22].

In a previous work [23] a linear extrapolation method for the demagnetizing field was used in a semi-implicit scheme. Here we address the problem of using polynomial extrapolation of the demagnetizing field for explicit evaluation methods in general. In particular, we show that when the polynomial approximation order is matched to the explicit evaluation method order, no, or negligible, increase in numerical error results. Indeed, higher order methods where the demagnetizing field is obtained through polynomial extrapolation at sub-steps are more accurate than lower order methods where the field is fully computed at all sub-steps. The advantage of this approach is a very significant reduction in computation time. We show it is possible to achieve speedup factors of 2 or more – largest speedup factor obtained was 2.5 – with no increase in numerical error.

## II. Method

Dynamical micromagnetic modelling uses the LLG equation to obtain the time evolution of magnetization:

$$\frac{\partial \mathbf{M}(\mathbf{r},t')}{\partial t} = -\gamma \mathbf{M} \times \mathbf{H}_{eff}(\mathbf{M}) + \alpha \frac{\mathbf{M}}{M_S} \times \frac{\partial \mathbf{M}}{\partial t} = \mathbf{F}(t', \mathbf{M}). \quad (1)$$

Here $\mathbf{M}$ is the magnetization vector, $M_S$ is the saturation magnetization, $\gamma$ is the gyromagnetic ratio, and $\alpha$ is the Gilbert damping constant. The effective field, $\mathbf{H}_{eff}$, includes a number of contributions, namely applied field contribution, the magnetostatic or demagnetizing field contribution, the direct exchange interaction, and magneto-crystalline anisotropy. Formulas for these interactions are well known and will not be repeated here – for example see Ref. [19]. The demagnetizing field in a cell-centred finite difference mesh may be obtained using a convolution sum:

$$\mathbf{H}(\mathbf{r}_0, t) = -\sum_{\mathbf{r}} \mathbf{N}(\mathbf{r} - \mathbf{r}_0) \mathbf{M}(\mathbf{r}, t). \quad (2)$$

Here $\mathbf{N}$ is the demagnetizing tensor, computed using formulas given in Ref. [24]. For the demagnetizing field in cell $\mathbf{r}_0$ the sum in Equation (2) runs over all the cells $\mathbf{r}$ in the discretization mesh. This convolution sum is efficiently evaluated using fast Fourier transforms, based on the convolution theorem, and constitutes the bulk of the computational effort in each iteration – see e.g. Ref. [25].

Explicit evaluation methods generally have the following form:

$$\begin{aligned} \mathbf{M} &= \mathbf{M}_0 + h \sum_k c_k \mathbf{F}_k + O(h^{n+1}) \\ \hat{\mathbf{M}} &= \mathbf{M}_0 + h \sum_k \hat{c}_k \mathbf{F}_k + O(h^{\hat{n}+1}), \end{aligned} \quad (3)$$

where

$$\mathbf{F}_k = \mathbf{F}\left(t_0 + a_k h, \mathbf{M}_0 + h \sum_{\lambda=0}^{k-1} b_{k\lambda} \mathbf{F}_\lambda \right), \quad (k = 1, \ldots, \max\{n, \hat{n}\} + 1). \quad (4)$$

Here $h$ is the time-step, $n$ is the evaluation method order, $\hat{\mathbf{M}}$ is magnetization evaluation of order $\hat{n}$ used to obtain the local truncation error for adaptive time-step control, $e = \|\mathbf{M} - \hat{\mathbf{M}}\| / M_S$. The coefficients $a_k$, $b_{k\lambda}$, $c_k$, and $\hat{c}_k$ depend on the evaluation method. In this work a number of explicit evaluation methods are used, both fixed time-step and adaptive time-step. In particular, for fixed time-step the 4$^{th}$ order Runge-Kutta (RK4) method is used [26]. Adaptive time-step methods are the adaptive Heun 2$^{nd}$ order, Runge-Kutta 3$^{rd}$ order with 2$^{nd}$ order error estimation (RK23), having first-same-as-last property [27]; Runge-Kutta-Fehlberg methods, 4$^{th}$ order with 5$^{th}$ order error estimation (RKF45) [28], respectively 5$^{th}$ order with 6$^{th}$ order error estimation (RKF56) [29]; Runge-Kutta-Cash-Karp 4$^{th}$ order with 5$^{th}$ order error estimation (RKCK45) [30]; Runge-Kutta-Dormand-Prince 5$^{th}$ order with 4$^{th}$ order error estimation (RKDP54) [31]; finally the 2$^{nd}$ order linear multi-step Adams-Bashforth-Moulton (ABM) predictor-corrector method [32] is also used. Adaptive time-step control is achieved using a standard integral controller [33]:

$$h' = \min\{\max\{hc, h_{\min}\}, h_{\max}\},$$

where

$$c = \min\left\{\max\left\{\left(\frac{0.8 e_{\max}}{e}\right)^{1/n+1}, c_{\min}\right\}, c_{\max}\right\}. \tag{5}$$

Here $h_{\min}$ and $h_{\max}$ are minimum and maximum time-step values, taken as 1 fs, respectively 3 ps. Bounds on the multiplicative constant $c$, namely $c_{\min}$ and $c_{\max}$ are taken as 0.01, respectively 2. The maximum local truncation error allowed, $e_{\max}$, is taken as $10^{-5}$. If the local truncation error $e$ exceeds this value the iteration is repeated with a smaller time-step.

When explicit evaluation methods are used to solve the LLG equation, all the effective fields – including the demagnetizing field – are conventionally computed at all sub-steps of the evaluation method in Equations (3) and (4). Here we show it is possible to use a much more efficient strategy for computing magnetization dynamics, with a negligible decrease in solution accuracy. This is based on evaluating the demagnetizing field at sub-steps using a cheap polynomial extrapolation method, thus requiring a single evaluation of Equation (2) per iteration. The general equation used in this work for polynomial extrapolation of order $p$ is given below:

$$\mathbf{H}(t) = \sum_{i=0}^{p} \left(\mathbf{H}_i - N_{00}\mathbf{M}_i\right) \prod_{\substack{j=0 \\ (j \neq i)}}^{p} \frac{t - t_j}{t_i - t_j} + N_{00}\mathbf{M}(t). \tag{6}$$

Here $t_i$ are start times of current and previous iterations, for which the magnetizations $\mathbf{M}_i$ and demagnetizing fields $\mathbf{H}_i$ – computed exactly from Equation (2) – are known. $N_{00}$ is the self-demagnetizing tensor element [24], and the stored fields are $\mathbf{H}_i - N_{00}\mathbf{M}_i$. The self-demagnetizing contribution, $N_{00}\mathbf{M}(t)$, is included explicitly at every sub-step without polynomial extrapolation, since it is computationally very cheap and forms the largest contribution to the demagnetizing field. Thus, using Equation (6) the demagnetizing field is approximated at sub-steps $t = t_0 + a_k h$ of the evaluation method, without requiring full computation of the expensive Equation (2) – the exception is for the first $p$ iterations which are required to prime the method. The order of the polynomial used is important as this affects the overall accuracy order of the evaluation method. At this point it is instructive to consider bounds on the approximation error. With the simplest method possible, where no extrapolation is used and the demagnetizing field is kept constant at sub-steps (zero order polynomial), from Equations (2) and (3) it is easy to see that $\|\Delta\mathbf{H}\| = \|\mathbf{H}(\mathbf{M}) - \mathbf{H}(\mathbf{M}_0)\| \leq hN \left\|\sum_k c_k \mathbf{F}_k\right\|$. Here $N$ is the number of cells in the finite-difference mesh. Thus the error is bounded by the time-step as expected, but also by the mesh size. This shows numerical tests of the method should also include large mesh sizes. With polynomial extrapolation of order $p$, in general the bound on the approximation error is of order $O(h^{p+1})$, thus we expect that a method of order $n$ should use a polynomial of at least order $p = n$ in order to achieve the same numerical error $h$ scaling.

## III. Results

In order to test the accuracy of the method discussed above, an approach similar to that used for μMAG Standard Problem #4 [34] is employed. Here a $Ni_{80}Fe_{20}$ slab is used (standard values $M_S$ = 800 kA/m, exchange stiffness $A$ = 13 pJ/m, no magneto-crystalline anisotropy, and $\alpha$ = 0.02), initialized with an S-state magnetization configuration. After initialization a reversal field is applied, and the average magnetization is recorded for 3 ns with a resolution of 1 ps. The main differences from μMAG Standard Problem #4 are the slab dimensions used, namely 3.2 μm × 1.6 μm × 20 nm, in order to allow for a greater computational effort and a large number of computational cells (over 800,000 with a 5 nm cubic cellsize), and the reversal field of (-20, 1, 0) kA/m. The simulated switching event is plotted in Figure 1(a), where the reference solution (RK4 with 10 fs time-step) is compared to that obtained using RKF56 with demagnetizing field quintic polynomial extrapolation, showing a virtually identical solution ($R^2$ > 0.999). Figure 1(b) shows the magnetization direction spatial variation obtained at $t$ = 0.5 ns using the reference RK4 method. This is also compared to the configurations obtained with all the evaluation methods, both with and without demagnetizing field extrapolation. The comparison is done using the normalized mean absolute error, Equation (7), where $\tilde{\mathbf{M}}_i$ are the points in the reference dataset containing $N$ points.

$$\varepsilon = \frac{1}{N} \sum_{i=1}^{N} \left\| \mathbf{M}_i - \tilde{\mathbf{M}}_i \right\| / M_S . \tag{7}$$

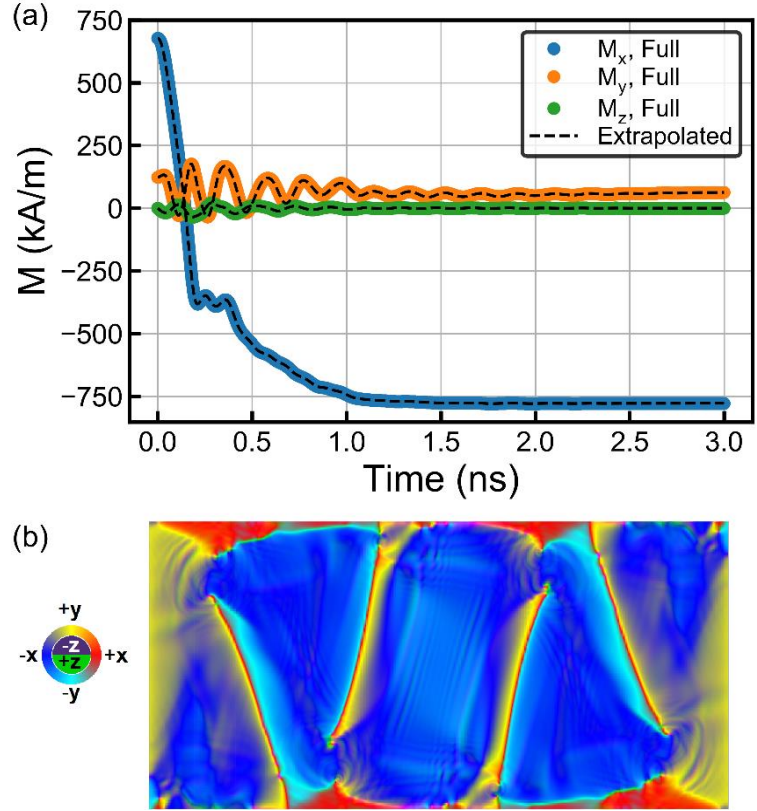

**Figure 1** – (a) Magnetization switching event in a 3.2 µm × 1.6 µm × 20 nm $Ni_{80}Fe_{20}$ slab, discretized using a 5 nm cubic cellsize, under an applied field of (-20, 1, 0) kA/m, showing the computed average magnetization components. The switching event is shown computed using the RK4 method with 10 fs time-step and full evaluation of demagnetizing field at all sub-steps (symbols), and using the RKF56 method with demagnetizing field obtained using quintic polynomial extrapolation at sub-steps (dashed lines). (b) Magnetization direction map during the switching event obtained at $t$ = 0.5 ns with RK4, with color-coded direction indicated by the color wheel.

First, the time-step error scaling of the methods is discussed. For this, µMAG Standard Problem #4 itself is used which shows a switching event similar to that in Figure 1 – see e.g. Ref. [19] for details. For each evaluation method a reference result is computed with a 10 fs time-step, and then the computation is repeated with fixed time-steps in the range 100 fs to 500 fs (adaptive time-step controller disabled). Results for the RKF56 method in double floating-point precision are shown in Figure 2(a). The computations are repeated using polynomial extrapolation with orders from $p$ = 1 (linear) up to $p$ = 5 (quintic), which may be compared to the results obtained with full evaluation of the demagnetizing field at all sub-steps. As expected, even though RKF56 is a 5$^{th}$ order method – $\varepsilon \propto h^5$ in Figure 2(a) – using polynomial extrapolation of order $p < 5$ limits the overall accuracy order of the method. With quintic extrapolation however the error values are very similar to the full computation results. This is repeated for all the evaluation methods, with results shown in Figure 2(b). Here, the full computation results are shown using solid symbols, and the results with polynomial extrapolation (polynomial order matched to the evaluation method order) are shown using open symbols. As may be seen, in all cases the errors are very similar; and we also notice that higher order methods with polynomial extrapolation are more accurate than lower order methods with full evaluation of demagnetizing field.

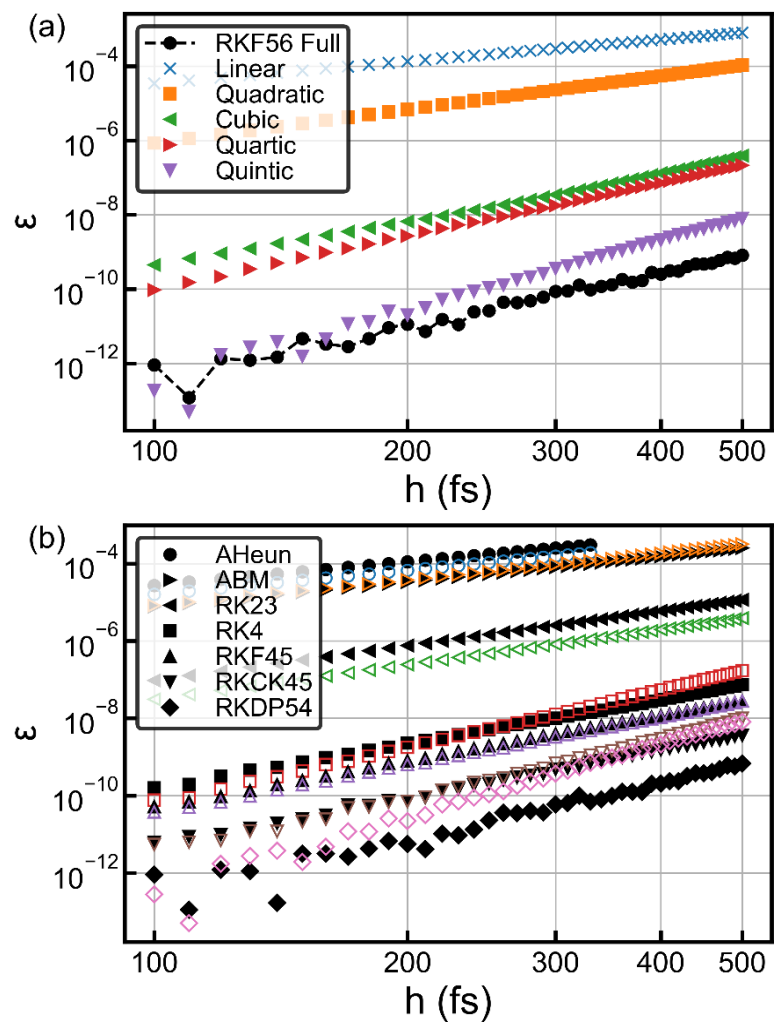

**Figure 2** – Normalized mean absolute errors, $\varepsilon$, as a function of time-step, $h$, for all the methods tested in double floating-point precision. (a) Results for RKF56 method with full evaluation of demagnetizing field, as well as polynomial extrapolation with polynomial orders varying from linear up to quintic. (b) Results for all other methods, where solid symbols are the evaluation methods with full evaluation of demagnetizing field, and open symbols are the respective results with polynomial extrapolation of order matched to the evaluation method order.

As a further test we also consider an array of nano-particles which interact only via the demagnetizing field. To this end a checkerboard nano-particle array is used, with mesh size comparable to that in μMAG Standard Problem #4, namely 300 nm × 300 nm × 5 nm, and with the same material parameters. The checkerboard array magnetization configuration is relaxed at zero field, which forms the starting state. A magnetic field of (20, 1, 0) kA/m is applied and the response also recorded over 3 ns with 1 ps resolution. Results are shown in Figure 3, where the same method for calculating $\varepsilon$ is used as for Figure 2. Again it is observed that the results with polynomial extrapolation (open symbols) are very similar to those with full calculation of the demagnetizing field (solid symbols). The only exceptions in this case are for the highest order methods (RKCK45, RKF56, and RKDP54) for time steps above 600 fs. It is unclear why this deviation arises, however the absolute errors are smaller than for RKF45 with exact demagnetizing field computation. It may be that a higher order polynomial extrapolation ($p > 5$) is required. This has not been tested, however the time steps used here are close to the data capture resolution (1 ps), which is an extreme use case.

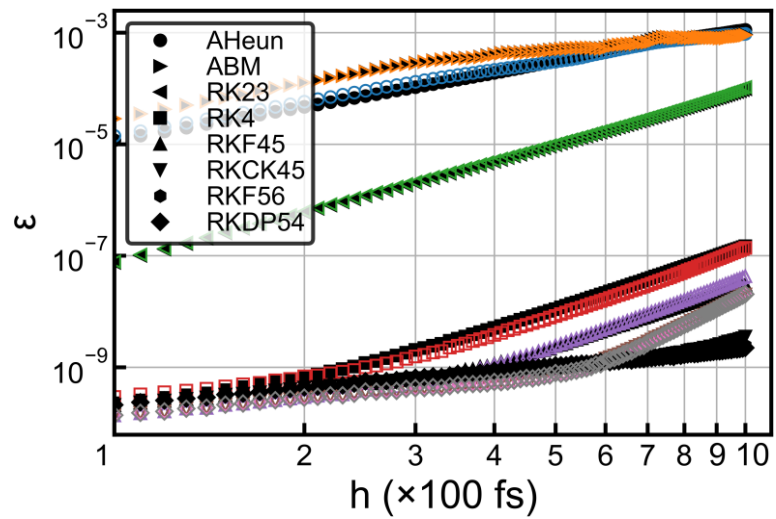

**Figure 3** – Normalized mean absolute errors, $\varepsilon$, as a function of time-step, $h$, for all the methods tested in double floating-point precision, using the checkerboard problem with no direct exchange contribution. Solid symbols show results with full computation of demagnetizing field, whilst open symbols are the respective results with polynomial extrapolation of order matched to the evaluation method order.

Next, the larger problem in Figure 1 is computed, but this time with the adaptive time-step controller enabled in order to simulate a real-case scenario. Also the results are now computed in single floating-point precision, as is common with LLG computations on the GPU.

As before, an accurate reference result is computed using RK4 with a small fixed time-step of 10 fs. Two different discretization schemes are used: 5 nm cubic cellsize and 2 nm cubic cellsize, where in each case the size of the mesh is limited to contain ~800,000 cells. It should be noted the exchange length in $Ni_{80}Fe_{20}$ is ~5 nm [35], which sets an upper limit for the discretization cellsize. It is well known that decreasing the cellsize is an important source of stiffness in the explicit evaluation of the LLG equation, requiring significantly smaller time-steps [36]. Since the demagnetizing field extrapolation method is expected to be more accurate for smaller time-steps, using the largest possible cellsize (where time-steps are maximized) is a particularly important test-case. The computed normalized mean absolute errors are shown in Figure 4. Looking first at the 2 nm discretization cellsize, the errors are virtually identical when comparing the full demagnetizing field computation with the polynomial extrapolation approach, for all the evaluation methods tested, and are also smaller than for the 5 nm discretization case as expected. The values obtained in this case are at the limit of detectable numerical error, limited by the single floating-point precision used. When looking at the 5 nm discretization cellsize, again comparing solutions with full demagnetizing field evaluation and polynomial extrapolation, the obtained errors are very similar. For all but one case (RK23) the errors obtained with polynomial extrapolation are smaller, however the change is negligible and cannot be distinguished from numerical noise. Moreover, the errors plotted in Figure 4 are all relatively small. Indeed, the mean absolute error obtained when comparing the reference datasets between single and double floating-point precision calculations is very similar to the values shown in Figure 4, namely $\varepsilon \cong 4\times10^{-4}$. These results put together show the polynomial extrapolation method for approximating the demagnetizing field at sub-steps is not a source of significant solution error. However, it is recommended as good practice to compare the two approaches for particular problem specifications before lengthy studies are conducted.

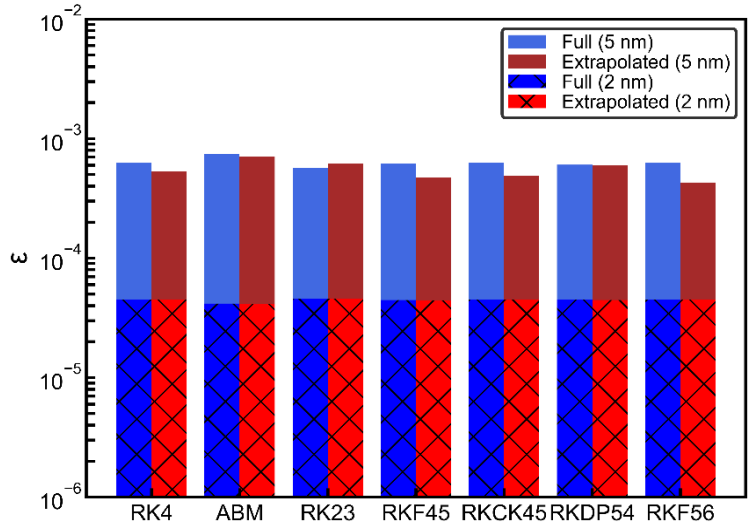

**Figure 4** – Normalized mean absolute errors, $\varepsilon$, shown for all the tested explicit evaluation methods, for the switching events computed with a 5 nm, respectively 2 nm cubic cellsize. The reference result was computed using the RK4 method with a 10 fs time-step in single floating-point precision on the GPU. The RK4 result shown here was computed using a 250 fs time-step for 5 nm cubic cellsize, respectively 100 fs for 2 nm cubic cellsize. For each evaluation method the left-hand-side bars represent the errors from simulations with full evaluation of the demagnetizing field, whilst the right-hand-side bars are obtained from simulations with polynomial extrapolation of the demagnetizing field. The results for 2 nm cubic cellsize are limited by the single floating-point precision.

Finally, the computational performance increase resulting from polynomial extrapolation is shown in Figure 5. Here, the physical computation time for the 5 nm discretization problem is shown in Figure 5(a). Very significant reductions in computation time are observed in all cases when the extrapolation method is used. For the fixed time-step RK4 method, the speedup obtained – ratio of computation time with full evaluation divided by computation time with polynomial extrapolation of the demagnetizing field – is to some extent predictable. Thus since RK4 requires 4 sub-step evaluations of the demagnetizing field for each iteration, which accounts for ~60 – 75% of the computation time, using a single evaluation of the demagnetizing field per iteration, with the much cheaper quartic extrapolation method applied at sub-steps, results in ~2 times faster computation time. For adaptive time-step methods the situation is slightly more complicated, since the time-step is variable and can also be affected by use of polynomial extrapolation rather than full evaluation of the demagnetizing field. However, in general it is to be expected that higher order methods result in larger speedup factors, and this is reflected in the results in Figure 5(b). Thus for example the ABM method, which only requires 2 sub-steps per iteration, shows the lowest speedup factors of ~1.5; similar remarks apply to RK23 which requires 3 sub-steps per iteration. This is to be compared to the RKF56 method which requires 8 sub-steps per iteration and shows large speedup factors of ~2.2 on average.

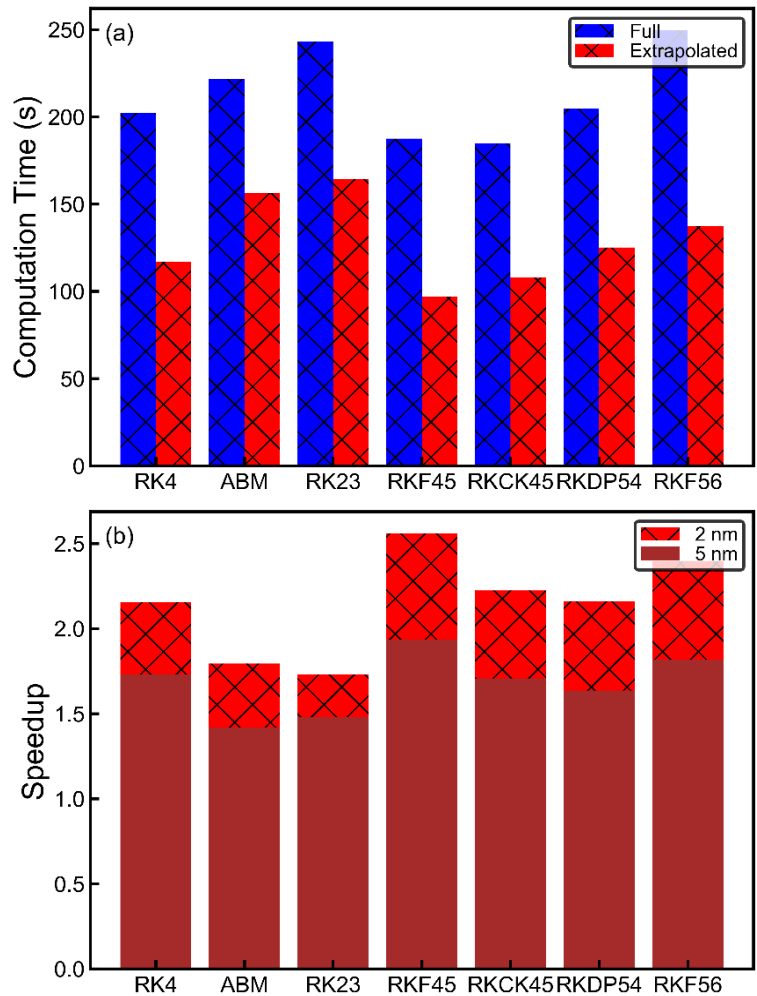

**Figure 5** – Increase in computational performance due to polynomial extrapolation of demagnetizing field for the tested explicit evaluation methods. (a) Computation times for the switching event simulation with 5 nm cubic cellsize. For each evaluation method the left-hand-side bars represent the computation time with full evaluation of the demagnetizing field, whilst the right-hand-side bars are obtained from simulations with polynomial extrapolation of the demagnetizing field. (b) Speedup factors obtained for the switching event computed with 5 nm, respectively 2 nm cubic cellsize, as computation time with full evaluation divided by computation time with polynomial extrapolation of the demagnetizing field.

# IV. Conclusions

A method of speeding up explicit evaluation methods for the LLG equation was presented here. This is based on approximating the demagnetizing field at sub-steps of higher order evaluation methods using polynomial extrapolation with self-demagnetizing contribution correction. A large number of explicit evaluation methods were tested, in common use in micromagnetic modelling software, both in single and double floating-point precision on the GPU. It was shown that polynomial extrapolation does not result in noticeable loss of simulation result accuracy. On the other hand the speedup factors obtained are up to 2 or more on average. In particular the RKF56 method was found to provide some of the largest speedup factors, whilst also exhibiting the smallest numerical errors. The fastest overall method was the RKF45, particularly when polynomial extrapolation is used. The accuracy required depends on the particular problem studied, and as expected higher order methods provide lower numerical errors. Moreover, it was shown that the use of polynomial extrapolation for the demagnetizing field does not significantly lower the numerical accuracy for the problems studied.

The implementation of this method is publically available in the Boris GitHub repository [37], which allows replication of this approach in other micromagnetic modelling implementations. The intended applications include magnetization dynamics simulations, for example magnetization switching of nanoparticles, thin films, spin valves and other multi-layered structures, domain wall motion studies, spin-torque-induced skyrmion dynamics, spin-wave, and spin-torque nano-oscillator simulations to name a few. For many such problems, which may involve lengthy parametric sweep studies, speedup factors of 2 or more are significant, particularly if no loss of accuracy arises.

# References

[1] T.L. Gilbert, "A phenomenological theory of damping in ferromagnetic materials" *IEEE Trans. Magn.* **40**, 3443 (2004).

[2] C. Thirion, W. Wernsdorfer, and D. Mailly, "Switching of magnetization by nonlinear resonance studied in single nanoparticles" *Nature Materials* **2**, 524-527 (2003).

[3] C.H. Back, R. Allenspach, W. Weber, S.S.P. Parkin, D. Weller, E.L. Garwin, and H.C. Siegmann, "Minimum Field Strength in Precessional Magnetization Reversal" *Science* **285**, 864-867 (1999).

[4] S. Kaka and S.E. Russek, "Precessional switching of submicrometer spin valves" *Appl. Phys. Lett.* **80**, 2958 (2002).

[5] B. Jinnai, C. Zhang, A. Kurenkov, M. Bersweiler, H. Sato, S. Fukami, and H. Ohno, "Spin-orbit torque induced magnetization switching in Co/Pt multilayers" *Appl. Phys. Lett.* **111**, 102402 (2017).

[6] T. Okuno, D.-H. Kim, S.-H. Oh, S.K. Kim, Y. Hirata, T. Nishimura, W.S. Ham, Y. Futakawa, H. Yoshikawa, A. Tsukamoto et al., "Spin-transfer torques for domain wall motion in antiferromagnetically coupled ferrimagnets" *Nature Electronics* **2**, 389-393 (2019).

[7] S. Lepadatu, M.C. Hickey, A. Potenza, H. Marchetto, T.R. Charlton, S. Langridge, S.S. Dhesi, and C.H. Marrows, "Experimental determination of spin-transfer torque nonadiabaticity parameter and spin polarization in permalloy" *Phys. Rev. B* **79**, 094402 (2009).

[8] S. Lepadatu, "Interaction of magnetization and heat dynamics for pulsed domain wall movement with Joule heating" *J. Appl. Phys.* **120**, 163908 (2016).

[9] K. Zeissler, S. Finizio, C. Barton, A. J. Huxtable, J. Massey, J. Raabe, A. V. Sadovnikov, S. A. Nikitov, R. Brearton, T. Hesjedal *et al.*, "Diameter-independent skyrmion Hall angle observed in chiral magnetic multilayers" *Nat. Commun.* **11**, 428 (2020).

[10] A. Hrabec, J. Sampaio, M. Belmeguenai, I. Gross, R. Weil, S. M. Chérif, A. Stashkevich, V. Jacques, A. Thiaville, and S. Rohart, "Current-induced skyrmion generation and dynamics in symmetric bilayers" *Nat. Commun.* **8**, 15765 (2017).

[11] C.R. MacKinnon, S. Lepadatu, T. Mercer, and P.R. Bissell, "Role of an additional interfacial spin-transfer torque for current-driven skyrmion dynamics in chiral magnetic layers" *Phys. Rev. B* **102**, 214408 (2020).

[12] K.G. Fripp and V.V. Kruglyak, "Spin-wave wells revisited: From wavelength conversion and Möbius modes to magnon valleytronics" *Phys. Rev. B* **103**, 184403 (2021).

[13] T. Taniguchi, "Synchronization and chaos in spin torque oscillator with two free layers" *AIP Advances* **10**, 015112 (2020).

[14] M.J. Donahue and D.G. Porter, "OOMMF User's Guide, Version 1.0" *Interagency Report NISTIR* **6376** (1999).

[15] A. Vansteenkiste, J. Leliaert, M. Dvornik, M. Helsen, F. Garcia-Sanchez, and B. Van Waeyenberge, "The design and verification of mumax3" *AIP advances* **4**, 107133 (2014).

[16] M.-A. Bisotti, D. Cortés-Ortuño, R. Pepper, W. Wang, M. Beg, T. Kluyver, and H. Fangohr, "Fidimag – A Finite Difference Atomistic and Micromagnetic Simulation Package" *Journal of Open Research Software*. 6, 22 (2018).

[17] C. Abert, G. Selke, B. Kruger, and A. Drews, "A Fast Finite-Difference Method for Micromagnetics Using the Magnetic Scalar Potential" *IEEE Trans. Magn.* **48**, 1105 (2012).

[18] R. Zhu, "The design, validation, and performance of Grace" *AIP Advances* **6**, 056401 (2016).

[19] S. Lepadatu, "Boris computational spintronics - High performance multi-mesh magnetic and spin transport modeling software" *J. Appl. Phys.* **128**, 243902 (2020).

[20] M. d'Aquino, C. Serpico, and G. Miano, "Geometrical integration of Landau–Lifshitz–Gilbert equation based on the mid-point rule" *Journal of Computational Physics* **209**, 730-753 (2005).

[21] D. Suess, V. Tsiantos, T. Schrefl, J. Fidler, W. Scholz, H. Forster, R. Dittrich, and J.J. Miles, "Time resolved micromagnetics using a preconditioned time integration method" *Journal of Magnetism and Magnetic Materials* **248**, 298-311 (2002).

[22] S. Fu, R. Chang, I. Volvach, M. Kuteifan, M. Menarini, and V. Lomakin, "Block Inverse Preconditioner for Implicit Time Integration in Finite Element Micromagnetic Solvers" *IEEE Trans. Mag.* **55**, 7205111 (2019).

[23] D. Shepherd, "Numerical methods for dynamic micromagnetics" *PhD Thesis*, University of Manchester, School of Computer Science (2015).

[24] A.J. Newell, W. Williams, and D.J. Dunlop, "A generalization of the demagnetizing tensor for nonuniform magnetization" *J. Geophys. Res. Solid Earth* **98**, 9551–9555 (1993).

[25] S. Lepadatu, "Efficient computation of demagnetizing fields for magnetic multilayers using multilayered convolution" *J. Appl. Phys.* **126**, 103903 (2019).

[26] E. Süli and M. David, "An Introduction to Numerical Analysis" *Cambridge University Press* (2003).

[27] P. Bogacki and L.F. Shampine, "A 3(2) pair of Runge-Kutta formulas" *Appl. Math. Lett.* **2**, 321-325 (1989).

[28] E. Fehlberg, "Low-order classical Runge-Kutta formulas with stepsize control and their application to some heat transfer problems" NASA Technical Report R-315 (1969).

[29] E. Fehlberg, "Classical fifth-, sixth-, seventh- and eight-order Runge-Kutta formulas with stepsize control" NASA Technical Report R-287 (1968).

[30] J.R. Cash and A.H. Karp, "A Variable Order Runge-Kutta Method for Initial Value Problems with Rapidly Varying Right-Hand Sides" *ACM Transactions on Mathematical Software* **16**, 201-222 (1990).

[31] J.R. Dormand and P.J. Prince, "A family of embedded Runge-Kutta formulae" *Journal of Computational and Applied Mathematics* **6**, 19-26 (1980).

[32] J.C. Butcher, "Numerical Methods for Ordinary Differential Equations" *John Wiley* (2013).

[33] K. Gustafsson, "Control theoretic techniques for stepsize selection in explicit Runge–Kutta methods" *ACM Transactions on Mathematical Software* **17** (1991).

[34] NIST micromagnetic modeling activity group (μMAG) website. https://www.ctcms.nist.gov/~rdm/std4/spec4.html. Accessed on July 7th, 2021.

[35] S. Lepadatu, "Effective field model of roughness in magnetic nano-structures" *Journal of Applied Physics* **118**, 243908 (2015).

[36] D. Shepherd, J. Miles, M. Heil, and M. Mihajlovic, "Discretization-induced stiffness in micromagnetic simulations" *IEEE Trans. Mag.* **50**, 7201304 (2014).

[37] Source code repository: https://github.com/SerbanL/Boris2. Accessed on July 7th, 2021.